%% file: PSCGTDBC_v1.tex
\documentclass[aps,pra,english,twocolumn,showpacs,preprintnumbers,amsmath,amssymb,floatfix,superscriptaddress,longbibliography]{revtex4-2}
\input{preamble}
\begin{document}
    \title{Giant-atom-enabled quantum optics with valley-polarized photons}
	
	\author{Marcel A. Pinto\,\orcidlink{0000-0002-8261-2283}}
	\affiliation{Universit$\grave{a}$ degli Studi di Palermo, Dipartimento di Fisica e Chimica -- Emilio Segr$\grave{e}$, via Archirafi 36, I-90123 Palermo, Italy}
	
\author{Giovanni Luca Sferrazza\,\orcidlink{0009-0006-7867-033X}}
\affiliation{Universit$\grave{a}$ degli Studi di Palermo, Dipartimento di Fisica e Chimica -- Emilio Segr$\grave{e}$, via Archirafi 36, I-90123 Palermo, Italy}
	
		\author{Silvia Casulleras\,\orcidlink{0000-0002-0009-759X}}
	\affiliation{Universit$\grave{a}$ degli Studi di Palermo, Dipartimento di Fisica e Chimica -- Emilio Segr$\grave{e}$, via Archirafi 36, I-90123 Palermo, Italy}

	\author{Alejandro~González-Tudela\,\orcidlink{0000-0003-2307-6967}}
	\affiliation{Quantum Advanced Research Center (QuARC), CSIC, Calle Serrano 113b, 28006 Madrid, Spain}
    \affiliation{Institute of Fundamental Physics (IFF), CSIC, Calle Serrano 113b, 28006 Madrid, Spain}

\author{Daniele De Bernardis\,\orcidlink{0000-0002-9618-0389}}
\affiliation{National Institute of Optics (CNR-INO), Via Nello Carrara 1, Sesto Fiorentino, 50019, Italy}
\affiliation{European Laboratory for Non-Linear Spectroscopy (LENS), Via Nello Carrara 1, Sesto F.no 50019, Italy}
	
	\author{Francesco Ciccarello\,\orcidlink{0000-0002-6061-1255}}
	\affiliation{Universit$\grave{a}$ degli Studi di Palermo, Dipartimento di Fisica e Chimica -- Emilio Segr$\grave{e}$, via Archirafi 36, I-90123 Palermo, Italy}
	
	\date{\today}
	
	\begin{abstract}

Valleytronics and valley photonics exploit the valley degree of freedom to encode and manipulate information. Here we show that photonic valleys can be selectively addressed in quantum optics using a simple two-level emitter, provided it is coupled nonlocally to the field, thereby realizing a so-called giant atom. Specifically, we consider a qubit coupled at multiple points to an engineered honeycomb lattice of resonators with detuned sublattice frequencies. By tailoring the geometry of the coupling points, the giant atom can be made to emit selectively into a single valley. The emitted photons thereby acquire a well-defined valley character and inherit the associated Berry curvature. By placing the qubit near a domain wall between regions of opposite sublattice detuning, whose interface supports valley-polarized edge modes, emission becomes chiral along the domain wall. 
This provides a promising route toward implementation of single-photon disorder-robust chiral emission without breaking time-reversal symmetry of the electromagnetic medium in platforms such as circuit QED.

	\end{abstract}
	
	
	\maketitle
	
	\section{Introduction}\label{sec-intro}

In condensed-matter physics, \emph{valleytronics} exploits the valley index of Bloch bands as an information carrier, analogous to spin in spintronics. In graphene and transition-metal dichalcogenides, inequivalent extrema at $K$ and $K'$ define a valley degree of freedom \cite{Xiao2007,Schaibley2016}; when inversion symmetry is broken, they acquire opposite Berry curvature and orbital magnetic moment, enabling valley-dependent transport and optical responses \cite{Xiao2007,Xu2014}, including optical valley polarization \cite{Mak2012} and the valley Hall effect \cite{Xiao2007,Gorbachev2014}.

An analogous valley-Hall mechanism exists in photonics: in engineered photonic crystals, inversion-symmetry breaking opens a topological band gap at the inequivalent $K$ and $K'$ valleys, which carry opposite valley character \cite{Ma2016AllSiValleyHall,Dong2017ValleyPhotonicCrystals}. This enables robust valley-selective transport through valley-Hall edge states, kink modes, and pseudospin--valley-coupled propagation \cite{Noh2018ObservationPhotonicTopologicalValleyHall,Gao2018TopologicallyProtectedRefraction,Kang2018PseudoSpinValleyCoupled}. Recent works have coupled quantum emitters to such valley-polarized edge modes \cite{Barik2020,Mehrabad2020,Mehrabad2023,Shi2023}, typically exploiting the correlation between valley character and local optical spin/helicity. This allows spin-selective coupling to opposite propagation channels, often using a V-type emitter with two circularly polarized transitions. Such schemes are appealing for quantum information processing, as they provide disorder-robust chiral single-photon emission \cite{Lodahl2017,SuarezForero2025}.

In platforms such as circuit QED \cite{Gu2017,Blais2021}, however, photons do not naturally provide a readily usable polarization degree of freedom. A different mechanism is therefore needed to selectively couple a quantum emitter to a given photonic valley.
	\begin{figure}[b]
	\centering
	   \includegraphics[width=0.85\linewidth]{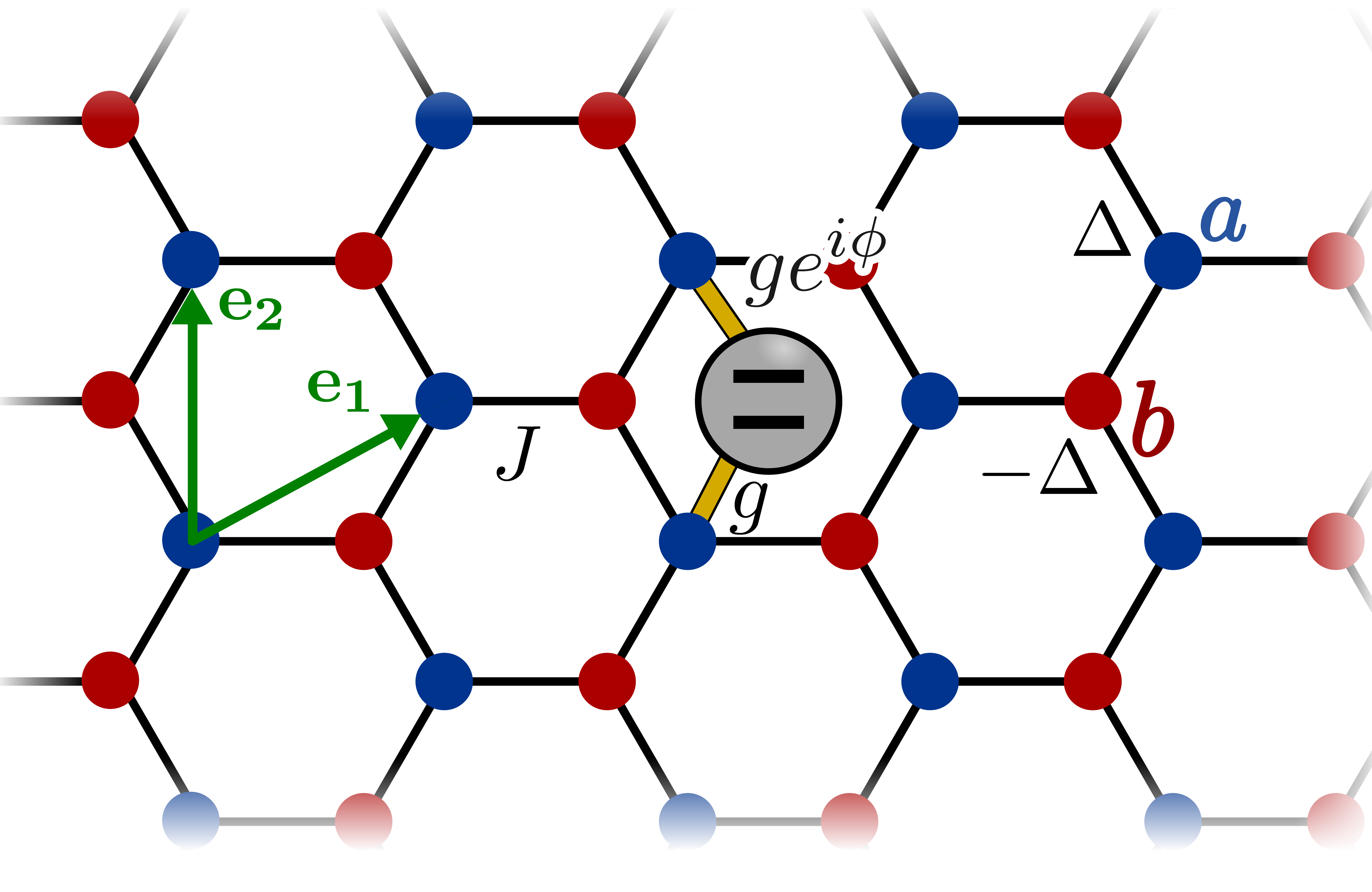}
	\caption{Sketch of the setup: a two-level giant atom is coupled to a photonic honeycomb lattice of resonators (blue and red dots). In this instance, the giant atom is coupled at two points, with coupling strengths $g$ and $g e^{i\phi}$, respectively. The bare lattice does not break time reversal symmetry, however it breaks inversion symmetry due to sublattice detuning measured by $\Delta$.}
	\label{fig:setup}
\end{figure}
In this work, we show that this task can be accomplished by using so-called {\it giant atoms} \cite{Kockum2021}. These have emerged in the last few years as a new paradigm in quantum optics, since they overcome the conventional description of an atom as a pointlike quantum emitter inherent in the electric-dipole approximation. In contrast to a small atom, a giant atom couples to the field at multiple discrete points, giving rise to a nonlocal light–matter interaction and, consequently, to a variety of distinctive phenomena \cite{Kockum2018-PRL, Kockum2014-PRA, Guo2017-PRA, wang2026decoherence, Levy-Yeyati-2026}.
Experimentally, giant atoms were first realized in circuit-QED platforms as superconducting qubits coupled to one-dimensional waveguides supporting either phonons or microwave photons \cite{Gustafsson2014,Kannan2020,Vadiraj2021}, and, more recently, using a ferromagnetic spin ensemble coupled to a meandering waveguide \cite{Wang2022}. Alternative implementations based on ultracold atoms and Rydberg atoms in photonic-crystal waveguides have been proposed \cite{GonzalezTudela2019,Chen2023}. In addition, the coupling at different points can be engineered to be complex and phase tunable, as recently demonstrated experimentally \cite{Joshi2023}. Very recently, giant-atom emission into an array of coupled resonators realizing the SSH model \cite{Su1979} was experimentally demonstrated in a circuit-QED setup by coupling a transmon qubit simultaneously to multiple resonators \cite{Jouanny2025}. 

A key property of giant atoms is their ability to enable {\it selective coupling} to photonic modes; for example, in a waveguide they can be engineered to couple to right-propagating modes while suppressing coupling to left-propagating ones \cite{Ramos2016,Joshi2023,Roccati-PRL-2024}. We exploit this nonlocal interference mechanism in a honeycomb photonic lattice \cite{GonzalezTudela2018,Perczel2020,RedondoYuste2021,DiBenedetto2025} to couple a giant atom to one valley while suppressing emission into the other. This valley selectivity imprints the Berry curvature of the selected valley onto the emitted photon. When a spatially varying sublattice detuning creates a topological domain wall, the same mechanism yields chiral single-photon emission along the interface.

Within the broad framework of waveguide QED \cite{Sheremet2023WQED,Ciccarello2024WQED}, and especially of circuit-QED arrays of coupled resonators interacting with qubits \cite{Liu2017,Mirhosseini2018,Sundaresan2019,Painter2021,Scigliuzzo2022,Owens2022,zhang2023,Hood2016,OBrien2025}, chiral emission from a quantum emitter locally coupled to a topological lattice has recently been explored experimentally in a circuit-QED implementation on a \(5\times 5\) lattice \cite{Owens2022}, as well as theoretically \cite{Vega2023TopologicalMultimode,DeBernardis2023ChiralBulk}. In all these cases, however, time-reversal symmetry is broken throughout the entire lattice, an experimentally demanding requirement for large systems. By contrast, our scheme achieves chiral emission while preserving time-reversal symmetry in the two-dimensional photonic lattice. The required symmetry breaking is confined to the light--matter coupling Hamiltonian and can be implemented with only two coupling points, one of which has a complex coupling amplitude. 

The remainder of the paper is organized as follows. We begin in Section \ref{sec-model} by introducing the model and reviewing the spectral and topological properties of the inversion-symmetry-broken honeycomb lattice. In Section \ref{sec-sel}, we show how the non-local coupling of a giant atom can be engineered to achieve selective coupling to a single valley. Section \ref{sec-SE} demonstrates that this mechanism yields valley-polarized spontaneous emission in the bulk. We then review in Section \ref{sec-edge} the valley-polarized edge modes supported by a topological domain wall. In Section \ref{sec-chiral}, we show that the giant atom can be made to couple only to the edge modes of a single valley, resulting in chiral single-photon emission along the domain wall. Finally, in Section \ref{sec-concl}, we summarize our results and discuss possible future directions.
	
	\section{Model and Hamiltonian} \label{sec-model}
	
	We consider a 2D honeycomb lattice of coupled single-mode resonators [see \ref{fig:setup}(a)], which embodies our photonic bath $B$, described by the Hamiltonian (throughout the text we set $\hbar=1$)
	\begin{equation}
		\label{eq:HB}
		H_B = \Delta \sum_{\R}(a_{\R}^\dagger a_{\R}- b_{\R}^{\dagger}b_{\R})+J \!\sum_{\expval{\R,\R'}} (a_{\R}^\dagger b_{\R'} + \Hc)\,.
	\end{equation}
Here, \(\mathbf{R}=n\mathbf{e}_1+m\mathbf{e}_2\) labels the unit cells, \(n=-N_1/2,\ldots,N_1/2-1\) and \(m=-N_y/2,\ldots,N_y/2-1\), where 
\begin{equation}\label{e1e2}
\mathbf{e}_1=\left(\frac{3a}{2}, \frac{\sqrt{3}a}{2}\right), \quad
\mathbf{e}_2=\left(0, \sqrt{3}a\right)
\end{equation}
are the primitive vectors, and $a$ is the nearest-neighbour distance \footnote{The notation distinguishes between $N_1$, the number of unit cells 
along the $\mathbf e_1$ direction, and $N_y$, the number of unit cells along 
the $\mathbf e_2$ direction. This choice is convenient because $\mathbf e_2$ 
is aligned with the $y$ axis.}. In the second sum, $\expval{\R,\R'}$ means that the sum runs over nearest-neighbour resonators, while $a_{\R}\equiv a_{n m}$ and $b_{\R}\equiv b_{n m}$ are standard bosonic ladder operators.  In \eq\eqref{eq:HB}, the parameter $\Delta$ is the \textit{sublattice detuning}, since $2\Delta$ is the difference between the bare frequencies of the $a$- and $b$-resonators associated with the two sublattices, while $J$ denotes the photon hopping rate [see \ref{fig:setup}]. For $\Delta=0$, \eq\eqref{eq:HB} is the bosonic analogue of the tight-binding Hamiltonian of pristine graphene. Importantly for the purposes of the present work, a nonzero detuning $\Delta$ breaks inversion symmetry in the lattice. We point out that, for any values of $J$ and $\Delta$, the free field Hamiltonian \eqref{eq:HB} does not break time-reversal symmetry. Still, the breaking of inversion symmetry for $\Delta\neq 0$ has some well-known topological consequences that will be reviewed in Section \ref{sec-berry}.

	A qubit (two-level quantum emitter), with ground (excited) state $\ket{g}$ ($\ket{e}$) whose energy separation is $\omega_0$, is {\it non-locally} coupled to the photonic lattice under the rotating-wave approximation. The total Hamiltonian thus reads
	\begin{equation}
		H= 	 H_B+\omega_0 \sigma^\dagger \sigma+H_{\rm int}	\label{Htot}
	\end{equation}
	with the atom-field coupling Hamiltonian given by
	\begin{equation}\label{Hint}
		H_{\rm int}=\frac{g}{\sqrt{\cal N}}\sum_{\ell=1}^{\cal N}\qty(e^{i \phi_\ell}  a_{\R_\ell}^{\dagger}\sigma+\Hc)\,,
	\end{equation}
	where $\sigma=\ketbra{g}{e}$ is the qubit pseudo-spin lowering operator, 
while $a_{\mathbf R_\ell}$ annihilates an excitation in the $a$-sublattice 
resonator located at position $\mathbf R_\ell$, with 
$\ell=1,\ldots,\mathcal N$. Each of these resonators is coupled to the qubit 
with strength $g_\ell=(g/\sqrt{\mathcal N})e^{i\phi_\ell}$, where 
$g/\sqrt{\mathcal N}$ is the common coupling modulus and $\phi_\ell$ is the 
corresponding phase. In the following, we refer to these resonators as 
``coupling points'' and to the qubit as a ``giant atom''. Notice that the 
coupling strengths $g_\ell$ are, in general, complex and may therefore break 
time-reversal symmetry. In Eq.~\eqref{Htot} and throughout the remainder of 
this work, we assume identical coupling magnitudes at all coupling points 
and restrict the giant atom to couple only to the $a$-resonators, which is 
sufficient for the purposes of this paper.
	
	We review next the derivation of the spectrum and normal modes of $H_B$.
	
	\subsection{Spectrum and normal modes of $H_B$}

\begin{figure}[b]
		\centering
		   \includegraphics[width=\linewidth]{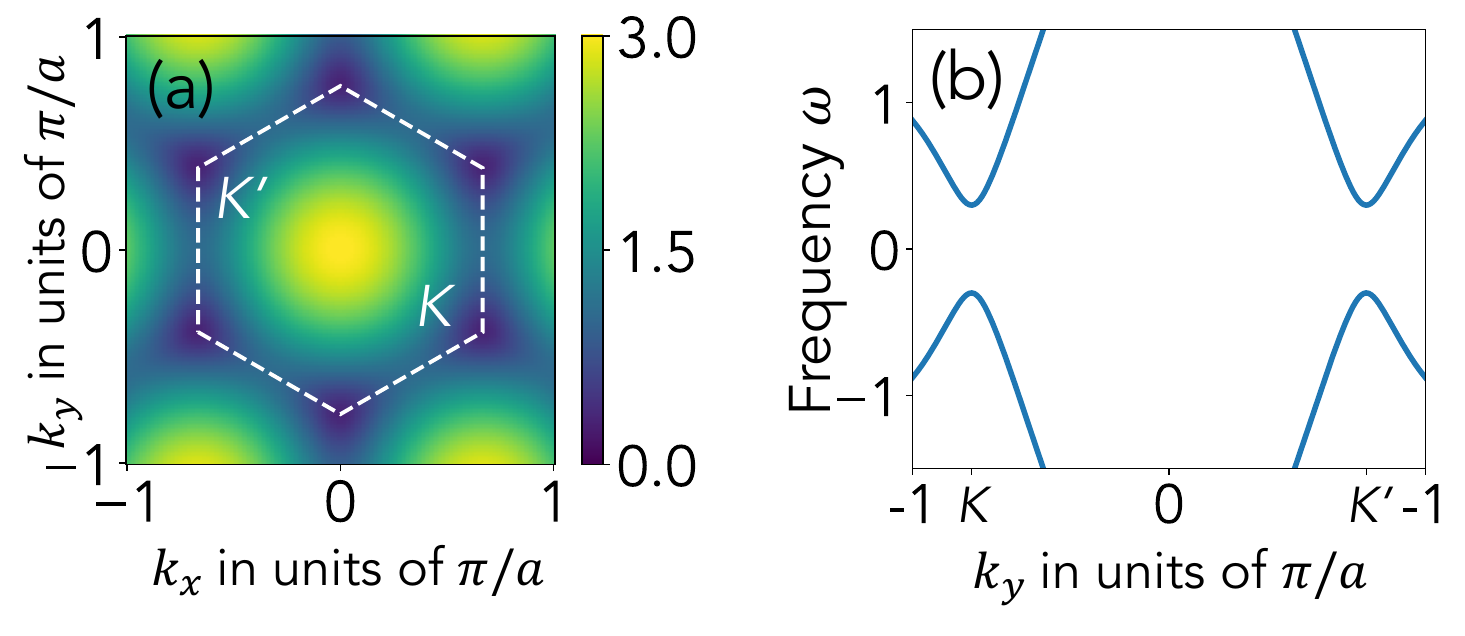}
		\caption{Bare lattice spectrum.
(a) Density plot of $\omega_{\mathbf{k}}$ [cf.\ Eqs.~\eqref{HBdiag} and \eqref{wk}]. 
The dashed line marks the boundary of the first Brillouin zone, whose two inequivalent vertices are the Dirac points $K$ and $K'$ (each of the four remaining vertices of the hexagon is equivalent to either $K$ or $K'$).
(b) Band dispersion along the $k_y$ direction for $k_x = 0$,
restricted to frequencies $|\omega| \le 5\Delta$ to highlight the dispersion near the \(K\) and \(K'\) valleys. 
In both panels, we set $\Delta=0.3J$ and frequencies are expressed in units of $J$.}
		\label{fig:FBZband}
	\end{figure}

	Under periodic boundary conditions, the lattice bare Hamiltonian $H_B$ enjoys discrete translational invariance with respect to the Bravais lattice ${\bf R}$. Accordingly, it is convenient to introduce the momentum ladder operators as
	\begin{equation}\label{akbk}
		a_{\mathbf k} = \frac{1}{\sqrt{N}} \sum_{\mathbf R} e^{-i \mathbf k \cdot \mathbf R} a_{\mathbf R}\,,
		\qquad
		b_{\mathbf k} = \frac{1}{\sqrt{N}} \sum_{\mathbf R} e^{-i \mathbf k \cdot \mathbf R} b_{\mathbf R}\,,
	\end{equation}
	where $N=N_1 N_y$ is the number of unit cells and $\mathbf{k}$ runs over the first Brillouin zone. This has hexagonal shape [see \ref{fig:FBZband}(a)] with vertices $K$ and $K'$ (Dirac points) of coordinates
	\begin{equation}\label{KK'}
		\mathbf{K} = -\mathbf{K}'=\left(\frac{2\pi}{3a},-\frac{2\pi}{3\sqrt{3}\,a}\right)\,.
	\end{equation}
	
	Upon inversion of \eqref{akbk}, Hamiltonian \eqref{eq:HB} can be arranged in the block-diagonal form (see e.g. \rrefs \cite{CastroNeto2009,FoaTorres2014})
	\begin{equation}
		H_B = \sum_{\mathbf k}
		\begin{pmatrix}
			a_{\mathbf k}^\dagger & b_{\mathbf k}^\dagger
		\end{pmatrix}
		h(\mathbf k)
		\begin{pmatrix}
			a_{\mathbf k} \\
			b_{\mathbf k}
		\end{pmatrix},
	\end{equation}
	where the $2\times 2$ Block Hamiltonian is defined as
	\begin{equation}\label{hk}
		h(\mathbf k) =
		\begin{pmatrix}
			\Delta & J f(\mathbf k) \\
			J f^*(\mathbf k) & -\Delta
		\end{pmatrix}
	\end{equation}
	with 
	\begin{equation}\label{fk}
		f(\mathbf k) {=}
		e^{i k_x a}
		{+} 2 e^{-i k_x a / 2}
		\cos\left(\frac{\sqrt{3}}{2} a k_y\right)\,.
	\end{equation}
	
	Through diagonalization of $h(\mathbf k) $, Hamiltonian $H_B$ can be eventually expressed in the diagonal form
	\begin{equation}\label{HBdiag}
		H_B = \sum_{\mathbf k}
		\omega_{\mathbf k}	\!\left( \alpha_{\mathbf k}^\dagger \alpha_{\mathbf k}
		- \beta_{\mathbf k}^\dagger \beta_{\mathbf k}\right)\,
	\end{equation}
	with 
	\begin{equation}\label{wk}
		\omega_{\mathbf k} = \sqrt{\Delta^2 + J^2 |f(\mathbf k)|^2}\,,
	\end{equation}
	and
	\begin{align}\label{alpha-beta}
		\alpha_{\mathbf k} &=
		\cos \frac{\theta_{\mathbf k}}{2}\, a_{\mathbf k}
		+
		e^{-i \varphi_{\mathbf k}} \sin \frac{\theta_{\mathbf k}}{2}\, b_{\mathbf k}, \\
		\beta_{\mathbf k} &=
		- e^{i \varphi_{\mathbf k}} \sin \frac{\theta_{\mathbf k}}{2}\, a_{\mathbf k}
		+
		\cos \frac{\theta_{\mathbf k}}{2}\, b_{\mathbf k}.
	\end{align}
	where 
	\begin{equation}\label{thetaphi}
\theta_{\mathbf k}={\rm arccos}(\Delta/\omega_{\mathbf k}),\,\,  \varphi_{\mathbf k}={\rm arg}\,f_{\bf k}.
	\end{equation}
	\\
	\\
	The spectrum thus consists of two symmetric bands with dispersion $\pm \omega_{\bf k}$ 
    separated by the energy gap $\Delta\omega= 2|\Delta|$ [see \ref{fig:FBZband}(a) and (b)]. Importantly, each band features a valley around each Dirac point [see \ref{fig:FBZband}(b)], which in the remainder we will refer to as valley $K$ and valley $K'$.
	
	\subsection{Valleys $K$ and $K'$}\label{sec-val}
	
	In each valley, sufficiently near the Dirac point, the band dispersion can be approximated as quadratic. Indeed, defining ${\bf K}_{+1} {=}{\bf K}$, ${\bf K}_{-1} {=}{\bf K}'{=}{-}{\bf K}$, we can approximate \eqref{fk} in valley $\tau=\pm 1$ as 
	\begin{equation}
		f(\mathbf{K}_{\tau}+\mathbf q)\simeq -\frac{3a}{2}(\tau q_y-i q_x)
	\end{equation}
	for $q\ll |{\bf K}_{\tau}|$, that is $q\ll 1/a$. Accordingly, in this regime, and upon an innocuous local redefinition of momentum coordinates, the Block Hamiltonian \eqref{hk} in each valley takes the form 
	\begin{equation}\label{htauq}
		h_\tau(\mathbf q)
		=
		v\left(\tau q_x \sigma_x + q_y \sigma_y\right)
		+
		\Delta \sigma_z\,,
	\end{equation}
	where we defined the characteristic speed
    \begin{equation}\label{speed}
       v=\frac{3}{2}aJ \,.
    \end{equation}
Hamiltonian \eqref{htauq} has eigenvalues $\pm \sqrt{\Delta^2+v^2q^2}$, in agreement with the lowest-order expansion of the band dispersion \eqref{wk} around each Dirac point.

Based on \eqref{htauq}, the effective total Hamiltonian in valley $\tau$ for low frequencies takes the form of a two-dimensional massive Dirac Hamiltonian \cite{CastroNeto2009,FoaTorres2014}
	\begin{equation}\label{htau}
	H_\tau
	=
	v\left(\tau p_x \sigma_x + p_y \sigma_y\right)
	+
	\Delta \sigma_z\,,
\end{equation} 
with $p_x$ ($p_y$) the effective momentum operator along the $x$ ($y$) axis.
	
	\subsection{Topological properties of valleys}\label{sec-berry}
	
	Based on the eigenstates of \eqref{htauq}, one can show that the Berry curvature \cite{GirvinYang2019} in each valley is given by \cite{XiaoRMP2010}
	\begin{equation}\label{berry-eq}
		{\bf \Omega}_\tau(\mathbf{q}) =  \tau\,\frac{v^2 \Delta}{2\left(\Delta^2+v^2 q^2\right)^{3/2}}\,\hat{z}\,.
	\end{equation}
	Therefore, the breaking of inversion symmetry ($\Delta\neq 0$) endows the two valleys with non-zero and opposite Berry curvatures. Also notice that ${\bf \Omega}_\tau$ is an odd function of $\Delta$. Thus, although the {\it total} Berry curvature $ {\bf \Omega}_{-}{+} {\bf \Omega}_{+}$ correctly integrates to zero as required by time-reversal symmetry \cite{GirvinYang2019}, the {\it valley-polarized} Berry curvature remains finite and gives rise to a nontrivial geometric phase for a particle adiabatically encircling a single Dirac valley $\vb{K}_{\tau}$. Accordingly, each  valley has a Chern number $C_{\tau}=\frac{1}{2}\tau\, \text{sgn}(\Delta)$. As $\Delta$ goes from a positive to a negative value, a valley-polarized phase transition takes place with the gap closing at $\Delta=0$.
	
	\section{Selective coupling to a single valley}\label{sec-sel}
	
	We are now interested in studying how a giant atom couples to the normal modes of the photonic honeycomb lattice. To this aim, we invert \eqs\eqref{akbk} and \eqref{alpha-beta} in order to express in terms of normal modes the real-space field operators appearing in the interaction Hamiltonian \eqref{Hint}. This yields
	\begin{equation}   \label{Hint-k}
	H_{\mathrm{int}}=
	\sum_{\mathbf{k}}
	(g_{\bf k}^+\,\alpha_{\mathbf{k}}^\dagger+g_{\bf k}^-\,\beta_{\mathbf{k}}^\dagger)\sigma
	+\mathrm{H.c.}\,,
	\end{equation}
	with the mode coupling functions for the upper and lower bands respectively given by
\begin{equation}
\begin{aligned}
g_{\bf k}^+&=\frac{g}{\sqrt{N \cal{N}}}\,\sum_{\ell=1}^{\cal N}
e^{i \phi_\ell}
e^{-i\mathbf{k}\cdot \mathbf{R}_\ell}
\cos\frac{\theta_{\mathbf{k}}}{2}\,,\\
g_{\bf k}^-&=-\frac{g}{\sqrt{N {\cal N}}} \sum_{\ell=1}^{\cal N}
e^{i \phi_\ell}
e^{-i\mathbf{k}\cdot \mathbf{R}_\ell}e^{i\varphi_{\mathbf{k}}}
\sin\frac{\theta_{\mathbf{k}}}{2}\,.
\end{aligned}
\end{equation}

In valley $\tau$, namely for $\mathbf k=\mathbf K_\tau+\mathbf q$ with $|\mathbf q|\ll |\mathbf K|$ [recall that $\tau{=}1$ ($\tau{=}{-}1$) corresponds to ${\bf K}$ (${\bf K}'$), with ${\bf K}'=-{\bf K}$], we have that for $\Delta>0$
\begin{equation}\label{g-pos}
g_{\mathbf K_\tau+\mathbf q}^{+}
\simeq
\frac{g}{\sqrt{N \cal{N}}}
\sum_{\ell=1}^{\mathcal N}
e^{i[\phi_\ell-(\tau {\mathbf K}+{\bf q})\cdot \mathbf R_\ell]},\,\,\,
g_{\mathbf K_\tau+\mathbf q}^{-}\simeq 0\,,
\end{equation}
while for $\Delta<0$
\begin{equation}\label{g-neg}
g_{\mathbf K_\tau+\mathbf q}^{+}\simeq 0,\,\,\,
g_{\mathbf K_\tau+\mathbf q}^{-}
\simeq
-\frac{g \,e^{i\varphi_{\tau,\mathbf q}} }{\sqrt{N {\cal N}}}
\sum_{\ell=1}^{\mathcal N}
e^{i[\phi_\ell-(\tau {\mathbf K}+{\bf q})\cdot \mathbf R_\ell]}\,.
\end{equation}
where  
$e^{i\varphi_{\tau,\mathbf q}}{=}(\tau q_x-iq_y)/q$.
Therefore, the giant atom couples predominantly to the upper (lower) band for $\Delta>0$ ($\Delta<0$). 
In either case, we see that the coupling to the valleys is governed by the same structure factor 
\begin{equation}\label{F-def}
F_\tau({\bf q})=\sum_{\ell=1}^{\mathcal N} e^{i[\phi_\ell-(\tau {\mathbf K}+{\bf q})\cdot \mathbf R_\ell]}\,.
\end{equation}
    
Now, since selective coupling to valley $\tau$ implies decoupling from the other, it is convenient to derive first the condition to decouple from a generic valley $\tau$. 

\subsubsection*{Decoupling condition from valley $\tau$}

Based on \eqref{g-pos} and \eqref{g-neg}, the condition to decouple from the whole valley $\tau$ reads $F_\tau({\bf q})\simeq 0$, that is
\begin{equation}
\sum_{\ell=1}^{\mathcal N} e^{i[\phi_\ell-(\tau {\mathbf K}+{\bf q})\cdot (\mathbf R_\ell-\mathbf R_1)]}\simeq 0\,\label{condK}
\end{equation}
for any ${\bf q}$ such that $q\ll 1/a$. 

In particular, this condition must hold for ${\bf q}=0$, corresponding to the valley center, which yields
\begin{equation}
\sum_{\ell=1}^{\mathcal N} e^{i[\phi_\ell-\tau {\mathbf K}\cdot (\mathbf R_\ell-{\bf R}_1)]}=0\,.
\label{condK-center}
\end{equation}
Arranging \eq\eqref{condK} in the form
\begin{equation}
\sum_{\ell=1}^{\mathcal N} e^{i[\phi_\ell-\tau {\mathbf K}\cdot ({\mathbf R}_\ell-{\mathbf R}_1)]}
\,e^{-i\mathbf q\cdot(\mathbf R_\ell-\mathbf R_1)}=0 
\end{equation}
we find that (approximate) decoupling from the entire valley occurs when 
\eqref{condK-center} holds together with $q\,|\mathbf R_\ell-\mathbf R_1|\ll 1$
for all coupling points and all momenta $\mathbf q$ in the considered valley. Since $q\ll 1/a$ within the valley, the latter condition corresponds to 
$|\mathbf R_\ell-\mathbf R_1|\sim a$.

\subsubsection*{Selective coupling condition}

Based on the above, we conclude that the conditions for the giant atom to couple selectively only to valley $\tau$ are thus
\begin{equation}\label{selcond-final}
\sum_{\ell=1}^{\mathcal N} z_{\ell}(\tau)\neq 0,\,\,\,\,\sum_{\ell=1}^{\mathcal N} z_{\ell}(-\tau)= 0,\,\,\,\,|\mathbf R_\ell-\mathbf R_1|\sim a\,,
\end{equation}
where for each coupling point we have defined the $\tau$-dependent phasor
\begin{equation}
z_\ell(\tau)=e^{i[\phi_\ell-\tau {\mathbf K}\cdot (\mathbf R_\ell-{\bf R}_1)]}\,.
\end{equation}
Geometrically, the second condition requires the ${\cal N}$ phasors $z_\ell(-\tau)$ to form a closed polygon in the complex plane, thereby ensuring decoupling from valley $-\tau$, whereas the corresponding polygon for valley $\tau$ must remain open (first condition), so that the coupling to that valley is nonzero. Finally, the third condition requires that the giant-atom size must be comparable to the unit-cell scale.

Notably, conditions \eqref{selcond-final} evidently cannot be satisfied by any normal atom (i.e., in the case ${\cal N}=1$), while they can be fulfilled by a giant atom (${\cal N}\ge 2$).

In particular, for {\it two coupling points}, the decoupling condition from valley $-\tau$ reduces to
\begin{equation}
1+e^{i[\phi+\tau \mathbf K\cdot (\mathbf R_2-\mathbf R_1)]}=0, \label{deca}
\end{equation}
where we set $\phi=\phi_2-\phi_1$. Hence
\begin{equation}\label{condK3}
\phi=\left[1-\frac{2\tau}{3}(\Delta n-\Delta m)\right]\pi\pmod{2\pi}
\end{equation}

with $\Delta n=n_2-n_1$ and $\Delta m=m_2-m_1$, where $\mathbf R_\ell=n_\ell \mathbf e_1+m_\ell \mathbf e_2$. It is easy to see that this also guarantees the first condition, namely a nonzero coupling to valley $\tau$, unless $\Delta n-\Delta m$ is an integer multiple of $3$. This possibility, however, is ruled out by the third condition in \eq\eqref{selcond-final}.

In particular, the condition can always be satisfied when the giant atom is coupled to a pair of $a$-resonators in nearest-neighbour cells (as in \ref{fig:setup}) since in this case either $\Delta n=\pm 1$, $\Delta m=0$ or $\Delta n=0$, $\Delta m=\pm 1$. In these cases, therefore, condition \eqref{selcond-final} yields either $\phi=\pi/3$ or $\phi=-\pi/3$ (up to a multiple integer of $2\pi$). In particular, in the instance of \ref{fig:setup} the coupling points are such that ${\bf R}_2{-}{\bf R}_1{=}\mathbf{e}_2$ hence $\Delta n=0$, $\Delta m=1$, implying that the giant atom couples only to valley $K$ ($K'$) for $\phi=\pi/3$ ($\phi=-\pi/3$).

	\begin{figure*}
	\centering
	   \includegraphics[width=\textwidth]{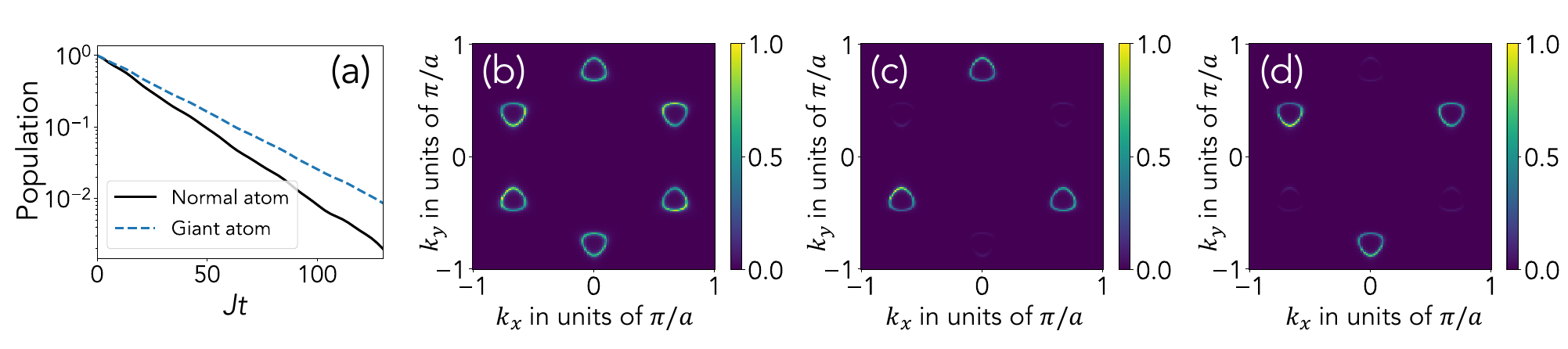}
	\caption{Homogeneous lattice: spontaneous emission of a normal and a giant atom for $\Delta=0.5J$, $\omega_0=0.7J$. The normal atom is coupled to resonator $a_{00}$ with strength $g=0.18J$, while the giant atom is coupled to $a_{00}$ and $a_{01}$ with strengths $g/\sqrt{2}$ and $g/\sqrt{2}\,e^{i \phi}$, respectively.  (a) Emitter's excited-state population $|\epsilon|^2$ versus time (in units of $J$) for the normal atom (black solid line) and the giant atom for $\phi=\pm \pi/3$ (blue dashed). (b) Momentum-space photon density $\rho({\bf k})=|\epsilon_{\alpha \mathbf k}|^2+|\epsilon_{\beta \mathbf k}|^2$ of the field emitted from the normal atom. (c) Same as (b), but for the giant atom with $\phi=\pi/3$. (d) Same as (c), but for $\phi=-\pi/3$. In (b)-(d), the photon density is computed at the time when $|\epsilon|^2=1\%$ and rescaled to its maximum value. All plots are obtained from numerical simulations of a lattice of $331\times 331$ unit cells.}
	\label{fig:test}
\end{figure*}
\subsubsection*{Weak-coupling regime}

It is worth noting that, in general, although the above conditions ensure coupling to only one valley, the giant atom may still couple to high-frequency lattice modes outside the two valleys. 
This possibility is, however, suppressed provided that such modes are far detuned from the emitter, which occurs when $\omega_0\approx 0$ is tuned near the band center, corresponding to the low-energy modes in proximity of the Dirac points, and the coupling strength is much smaller than the hopping rate $g\ll J$. 
This condition is completely consistent also with the previous decoupling condition from the $-\tau$ valley. 
Indeed, by considering Eq. \eqref{htau}, we require that the coupling is not larger than the maximum energy consistent with the Dirac Hamiltonian approximation, i.e. $g\ll v \,q_{\rm max}$, for a wavevector $0<q_{\rm max}<\pi/a$. From Eq. \eqref{speed}, as an upper bound, we immediately get $g/J\ll 3/2$.
In the following, we will always consider this weak-coupling regime.

\section {Spontaneous emission of valley-polarized photons}\label{sec-SE}

To assess the effectiveness of selective coupling to a single valley, we compare the spontaneous emission of a normal and a giant atom in the bulk of a homogeneous honeycomb lattice according to Hamiltonian \eqref{Htot} (we will consider a non-homogeneous lattice later on). The joint state at time $t$ lies in the single-excitation sector and has the form 
\begin{equation}
	\ket{\Psi(t)} = \epsilon(t)\ket{e}+ \sum_{\mathbf k} \epsilon_{\alpha \mathbf k}(t)\,
	\ket{\alpha_ {\bf k}} +\sum_{\mathbf k} \epsilon_{\beta \mathbf k}(t)
	\ket{\beta_ {\bf k}} ,
	\label{psit}
\end{equation}
where $ \ket{e}$ is the state where the emitter is in the excited state and the field in the vacuum state $\ket{\rm vac}$, while $\ket{\alpha_ {\bf k}}=\ket{g}\alpha_ {\bf k}^\dag \ket{\rm vac}$ and $\ket{\beta_ {\bf k}}=\ket{g}\beta_ {\bf k}^\dag \ket{\rm vac}$. The initial condition is $\epsilon(0)=1$, $\alpha_{\mathbf k}(0)=\beta_{\mathbf k}(0)=0$. Replacing \eqref{psit} into the Schr\"odinger equation $\frac{d}{dt}\ket{\Psi(t)}=-iH \ket{\Psi(t)}$ leads to a linear first-order differential system in the $2N+1$ unknown functions $\epsilon(t)$, $\alpha_ {\mathbf k}(t)$ and $\beta_{\mathbf k}(t)$, which can be easily solved.

We consider first a normal atom coupled to resonator $a_{00}$ with strength $g$ and then a giant atom coupled to $a_{00}$ and $a_{01}$ with strengths $g/\sqrt{2}$ and $g/\sqrt{2}\,e^{i \phi}$, respectively [\cf\eq\eqref{Hint}].
In either case, we set $\Delta >0$, $\omega_0>\Delta$ and $g$ small compared to $\omega_0-\Delta$ corresponding to an emitter weakly coupled to the upper photonic band and far-detuned from the lower band (hence the latter is negligibly populated) 
\footnote{Setting a non-zero $\Delta$ is not necessary for the essential qualitative results of the present section, which would hold even for $\Delta=0$ provided that $\omega_0$ is sufficiently far from the Dirac points.}. The resulting emission dynamics is displayed in \ref{fig:test}. As expected, both the normal and giant atoms undergo a standard exponential decay although with different rates [see \ref{fig:test}(a)]. In momentum space, the field emitted from the normal atom [see \ref{fig:test}(b)] is uniformly distributed along thin contours around the points ${\bf K}$ and ${\bf K}'$ (as well as the remaining four vertices of the first Brillouin zone, which are equivalent to ${\bf K}$ or ${\bf K}'$). Remarkably, the contour around the point $\mathbf{K}'$ (as well as those around the two equivalent points) vanishes for the field emitted by the giant atom when $\phi=\pi/3$ [see \ref{fig:test}(c)]. Setting instead $\phi=-\pi/3$ conversely suppresses the contour around the point $\mathbf{K}$ [see \ref{fig:test}(d)]. 

The behavior of the emitted field in panels (c) and (d) confirms the validity of the condition for selective coupling to a specific valley derived in the previous section.
This shows that a giant atom with suitably engineered coupling points (even just two) can emit valley-polarized light. 

It is worth noting that, as is well known, while being preserved globally, time-reversal symmetry is broken within {\it each} valley. This is consistent with the condition \eqref{condK3} implying that valley-polarized emission requires the atom–field coupling Hamiltonian \eqref{Hint} to break time-reversal symmetry.

\section{Valley-polarized edge modes} \label{sec-edge}

As discussed in Section~\ref{sec-berry}, breaking inversion symmetry through a non-zero detuning $\Delta$ between sublattice frequencies endows each valley with a nontrivial topology, as signaled by a nonzero Berry curvature whose sign is controlled by $\Delta$ [cf.~Eq.~\eqref{berry-eq}]. A remarkable consequence of this valley topology is that, when two lattice domains with opposite sublattice detuning ($\Delta=\pm \Delta_0$) are engineered, within each valley a topological phase transition occurs across the interface. As a result, a band of gapless, {\it valley-polarized edge modes} emerges along the domain wall \cite{Jun11,Noh17}. In the present section, we review the properties of these edge modes.

In the presence of a spatially-varying detuning $\Delta$, the lattice Hamiltonian \eqref{eq:HB} becomes
\begin{equation}
	\label{eq:HB2}
	H_B = \sum_{\R}  \Delta({\bf R})(a_{\R}^\dagger a_{\R}- b_{\R}^{\dagger}b_{\R})+J \!\sum_{\expval{\R,\R'}} (a_{\R}^\dagger b_{\R'} + \Hc)\,,
\end{equation}
where $\Delta({\bf R})$ changes sign upon crossing a domain wall. A standard and convenient way to model $\Delta({\bf R})$ so as to ensure a smooth spatial sign change across the domain wall is 
\begin{equation}\label{DeltaR}
	\Delta(\mathbf{R}) = \Delta_0 \tanh\!\left(\frac{\mathbf{R}\cdot \hat{\mathbf{v}}}{\lambda}\right),
\end{equation}
which interpolates between $\pm \Delta_0$ over a length scale $\lambda$, where $\hat{\mathbf{v}}$ identifies the gradient direction. We assume $\lambda \gg a$, such that the detuning varies slowly on the scale of the lattice spacing.

In the following, we set $\hat{\mathbf{v}}=\hat{\mathbf{x}}$, so that the domain wall is centered at $x=0$ and extends along the $y$ direction; the results, however, are naturally generalized to an arbitrary $\hat{\mathbf{v}}$.

\subsection{Edge modes}

	\begin{figure}[b]
	\centering
	  \includegraphics[width=\linewidth]{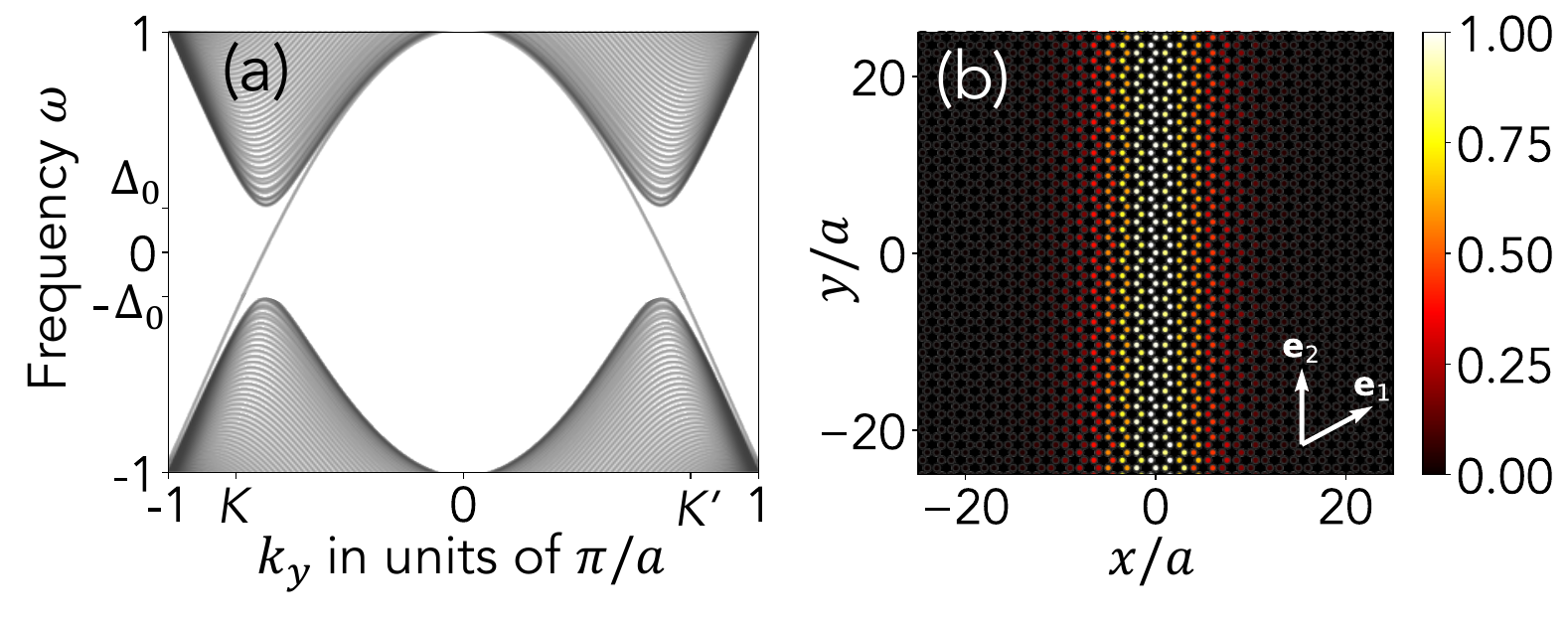}
	\caption{Valley-polarized edge modes under a spatially varying sublattice detuning $\Delta(x)=\Delta_0{\rm tanh}  (x/\lambda)$ for $\Delta_0 = 0.2 J$, and $\lambda = 4 a$.
(a) Lattice dispersion along $k_y$ (frequency is in units of $J$). 
(b) Rescaled real-space photon density of the lattice normal mode at frequency $\omega = -0.16J$ for $k_y = -0.73\pi/a$.
The plots were obtained by numerical diagonalization of $H_B(k_y)$, the partial Fourier transform of \eqref{eq:HB2} along the $y$ axis, for $k_y$ running over the first Brillouin zone, using a lattice of $100\times 100$ unit cells. 
}
	\label{fig:edge}
\end{figure}
In \ref{fig:edge}, we show the results of numerically diagonalizing \eqref{eq:HB2} after Fourier transformation along the $y$ direction. As expected, two band dispersions  with opposite slopes arise  within the gap $|\omega|{\le }\Delta_0$ around the $K'$ and $K$ points, respectively [see \ref{fig:edge}(a)]. Each in-gap band corresponds to valley-polarized edge modes propagating along the $y$ axis and localized across the domain wall at $x=0$, as shown in \ref{fig:edge}(b) for a representative edge mode. Notably, the edge modes associated with the two valleys have opposite group velocities.

Using the standard envelope-function approximation \cite{Asboth2016ShortCourse} and replacing \(\Delta \to \Delta(x)\) in the effective massive Dirac Hamiltonian \eqref{htau} for each valley \(\tau\), one can derive analytical expressions for the edge modes and their spectrum near \(\omega=0\), where the dispersion is approximately linear [see \ref{fig:edge}(a)].

The resulting edge modes can be written in real space as single-photon Fock states 
\begin{equation}\label{edge-state}
\ket{{\cal E}_{\tau,q_y}}
=
\sum_{\mathbf R}
\left[
\Psi_{\tau,q_y}(\mathbf R)\,a^\dagger_{\mathbf R}
+
\Phi_{\tau,q_y}(\mathbf R)\,b^\dagger_{\mathbf R}
\right]
\ket{\rm vac},
\end{equation}
with (see Appendix \ref{appA})
\begin{equation}
\Psi_{\tau,q_y}(\mathbf{R})
=-i\tau\,\Phi_{\tau,q_y}(\mathbf R)
=
e^{i\tau \mathbf K\cdot \mathbf R}
\frac{e^{iq_y y_{\mathbf R}}}{\sqrt{N_y}}\,
\frac{\psi(x_{\mathbf R})}{\sqrt{2}}\,,
\label{Psi-tau}
\end{equation}
where \(q_y=\nu 2\pi/L_y\) with $L_y=N_y\sqrt{3}\,a$ the system length along \(y\) [cf. Eq.~\eqref{e1e2}], while \(x_{\mathbf R}\) and \(y_{\mathbf R}\) are the Cartesian components of \(\mathbf R\). The transverse localized wavefunction appearing in Eq.~\eqref{Psi-tau} is defined as
\begin{equation}\label{psi-bound}
\psi(x_{\bf R})=A\exp\!\left[-\frac{1}{v}\int_0^{x_{\bf R}}\Delta(x')\,dx'\right],
\end{equation}
where $A$ is fixed by the normalization of \eqref{edge-state} and is explicitly given by
\begin{equation}\label{A-norm}
A=
\left[
\sum_{n=-N_x/2}^{N_x/2-1}
\exp\!\left(
-\frac{2}{v}\int_0^{\tfrac{3a}{2}n}\Delta(x')\,dx'
\right)
\right]^{-1/2},
\end{equation}
where we used \(x_{\mathbf R}=\tfrac{3a}{2}n\) [see Eq.~\eqref{e1e2}].

The corresponding spectrum of Eq.~\eqref{edge-state} reads
\begin{equation}\label{Etauq}
\omega_\tau(q_y)=\tau v q_y \, .
\end{equation}
with $v$ given by \eq\eqref{speed}.

Therefore, the edge modes propagate along \(y\) with group velocity \(+v\) (\(-v\)) in the \(K\) (\(K'\)) valley and are localized along \(x\), with an asymptotic decay \(\sim e^{-|x|/\xi}\), where
\begin{equation}\label{eq-xi}
\xi=\frac{v}{\Delta_0}\,,
\end{equation}
thus setting their transverse width [we used $\Delta(x){\to} \pm \Delta_0$ for $ x{\to} \pm\infty$].

In the specific case of the detuning profile $\Delta(x)=\Delta_0 \tanh(x/\lambda)$ [\cf\eq \eqref{DeltaR}], the transverse wavefunction \eqref{psi-bound} explicitly reads (see Appendix~\ref{appA})
\begin{equation}\label{psix}
\psi(x_{\bf R})=\left[\frac{1}{\lambda\sqrt{\pi}}
\frac{\Gamma\!\left(\beta+\tfrac{1}{2}\right)}
{\Gamma(\beta)}\right]^{1/2}\!
\cosh^{-\beta}\!\left(\frac{x_{\bf R}}{\lambda}\right),
\end{equation}
where \(\beta=\lambda/\xi\). For large \(|x_{\bf R}|\), \(\psi(x_{\bf R})\) decays asymptotically as \(e^{-|x_{\bf R}|/\xi}\).

\section {Single-photon chiral emission} \label{sec-chiral}

In the presence of $\Delta(x)$ and in the weak-coupling regime, a quantum emitter tuned within the band gap couples only to the valley-polarized edge modes, as the bulk modes are far detuned and can be neglected. In this section, we first analyze this spontaneous emission process for a normal atom and then generalize it to a giant atom.

\subsection{Emission from a normal atom}

\begin{figure*}
	\centering
	  \includegraphics[width=\textwidth]{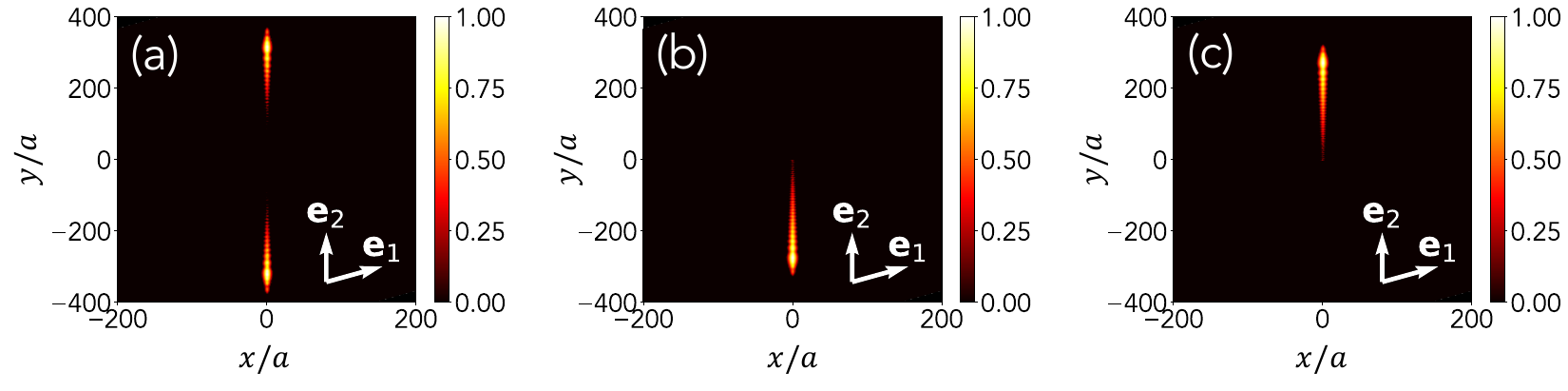}
	\caption{Lattice with a non-uniform sublattice detuning $\Delta(x)=\Delta_0\, {\rm tanh}(x/\lambda)$: spontaneous emission of a normal and a giant atom for $\Delta_0=0.5J$, $\lambda=2a$,  $\omega_0=0$. The normal atom is coupled to resonator $a_{00}$ (lying at the origin) with strength $g=0.3J$, while the giant atom is coupled to $a_{00}$ and $a_{01}$ with strengths $g/\sqrt{2}$ and $g/\sqrt{2}\,e^{i \phi}$, respectively.  (a) Log-scale rescaled real-space photon density of the field emitted from the normal atom. (b) Same as (a), but for the giant atom setting $\phi=-\pi/3$. (c) Same as (b) but for $\phi=\pi/3$. In all cases, we numerically verified that the emitter decays exponentially, with decay rates $\Gamma =\sum_\tau \Gamma_\tau= 0.03J$ in (a) and, $\tilde\Gamma_{+ 1} = 0.021J$ in (b) and $\tilde\Gamma_{- 1} = 0.021J$ in (c), in excellent agreement with \eqs\eqref{Gamma-tau} and \eqref{Gamma-giant}. In all plots, the photon density is computed at the time when $|\epsilon|^2=1\%$.  All plots are obtained from numerical simulations of a lattice of $551\times 551$ unit cells.}
	\label{fig:chiral}
\end{figure*}

Consider a normal atom tuned near the band gap center [\cf\eq\eqref{Hint} for ${\cal N}=1$] and coupling strength $g$ small compared to $\Delta_0$ \footnote{We place the emitter at the center of the band gap for the sake of argument. The essential requirement is that its frequency lies within the gap and its coupling is small compared to the detuning from band edges.}.
\ref{fig:chiral}(a) shows the real-space photon density of the field emitted from an atom coupled to the resonator $a_{{\bf R}=0}$. The emitted photon clearly propagates along the domain wall, equally in the upward and downward directions
thus providing evidence that the normal atom couples with the same strength to the edge modes of both valleys. This conclusion can be quantitatively supported by explicitly calculating the decay rate into each valley, as outlined below.

Let $\rho_\tau(\mathbf{R}; \omega)$ denote the local density of states of the valley-polarized edge modes at the $a$-resonator in unit cell $\mathbf{R}$. Based on \eqs\eqref{Psi-tau} and \eqref{Etauq}, this is given by
\begin{align}\label{DOS}
\rho_\tau(\mathbf{R}; \omega)&=\sum_{q_y}|\Psi_{\tau,q_y}(\mathbf{R})|^2  \delta\!\left[\omega-\omega_\tau(q_y)\right]\nonumber\\
&=\frac{1}{2N_y}\sum_{q_y} |\psi(x_{\bf R})|^2 \delta (\omega-\tau v q_y) \nonumber\\
&\simeq\frac{1}{2N_y}\frac{L_y}{2\pi v}|\psi(x_{\bf R})|^2 \nonumber\\
&=
\frac{\sqrt{3}\,a}{4\pi v}\,|\psi(x_{\mathbf R})|^2\,.
\end{align}
This expression is independent of $\omega$, owing to the linear dispersion in Eq.~\eqref{Etauq}, and also of $\tau$, since this appears only in the phases of the exponential factors [\cf\eq\eqref{Psi-tau}]. 
Accordingly, the decay rate into valley $\tau$ for a normal atom coupled to the $a$-resonator in the unit cell $\mathbf{R}$ is given by
\begin{align}\label{Gamma-tau}
\Gamma_\tau({\bf R})&=2\pi g^2  \rho_\tau(\mathbf{R};\omega_0)\nonumber\\
&=\frac{g^2}{\sqrt{3}\,J}\,|\psi(x_{\mathbf R})|^2\,,
\end{align}
where we used \eq\eqref{speed}.
Since the right-hand side of \eqref{Gamma-tau} is independent of $\tau$, the normal atom decays into both valleys at equal rates with the total decay rate given by $\Gamma=\sum_\tau \Gamma_\tau$.
 
An explicit expression for $\Gamma_\tau$ can be obtained by substituting \eq \eqref{psix} into \eq\eqref{Gamma-tau}, which in particular allows one to predict the numerically computed decay rate in the case of \ref{fig:chiral}(a).

\subsection{Emission from a giant atom}

We now replace the normal atom with a giant atom coupled to resonators $a_{00}$ and $a_{01}$ with relative phase $e^{i\phi}$ with $\phi=\phi_2-\phi_1$. \ref{fig:chiral}(b) shows the real-space photon density emitted by the giant atom for $\phi=-\pi/3$. While the photon still propagates along the domain wall, it now does so exclusively in the downward direction, thus demonstrating chiral emission. For $\phi=\pi/3$, as shown in \ref{fig:chiral}(c), the propagation is instead fully upward. This shows that the giant atom selectively couples to only one band of valley-polarized edge modes, with emission into the other band being fully suppressed.

For a giant atom, the decay rate into valley $\tau$ does not depend on the standard local density of states \eqref{DOS}, but is instead given by \cite{Leonforte2025}
\begin{equation}\label{Gamma-giant1}
\tilde\Gamma_\tau=2\pi g^2 \tilde{\rho}_\tau(\omega_0)
\end{equation}

with $\tilde{\rho}_\tau$ an effective local density of states defined by [\cf\eq\eqref{Hint}]

\begin{align}
\tilde\rho_\tau(\omega)&=\sum_{q_y}\frac{1}{\cal N}\left|\sum_{\ell=1}^{\cal N} e^{-i\phi_\ell}\Psi_{\tau,q_y}(\mathbf{R}_\ell)\right|^2  \delta\!\left[\omega-\omega_\tau(q_y)\right]\,.\label{chi-def}
\end{align}
Explicitly, using \eqs\eqref{Psi-tau} and \eq\eqref{Etauq}
\begin{align}\label{Gamma-giant3}
\tilde \rho_\tau(\omega)
&=
\frac{1}{2N_y }
\sum_{q_y}
\frac{1}{\cal N}\left|
\sum_{\ell=1}^{\mathcal N}
\psi(x_{\mathbf R_\ell})
e^{\,i\left(
\tau \mathbf K\cdot \mathbf R_\ell
+ q_y y_{\mathbf R_\ell}
-\phi_\ell
\right)}
\right|^2
\nonumber\\
&\qquad\times
\delta(\omega-\tau v q_y)\,.
\end{align}
For a giant atom of size comparable with the unit cell and for $\xi\gg a$, that is $\Delta_0\ll J$ [\cf\eqs\eqref{speed} and \eqref{eq-xi},  we can approximate $\psi(x_{{\bf R}_\ell})\simeq \psi(x_{{\bf R}_1})$ and  $q_y (y_{{\bf R}_\ell}-y_{{\bf R}_1})\simeq 0$, hence 
\begin{equation}\label{Gamma-giant4}
\tilde \rho_\tau(\omega)\simeq
\frac{\sqrt{3}\,a}{4\pi v}\,
|\psi(x_{{\bf R}_1})|^2
\frac{1}{\cal N}\left|
\sum_{\ell=1}^{\mathcal N}
e^{i(\tau {\bf K}\cdot {\bf R}_\ell-\phi_\ell)}
\right|^2\,.
\end{equation}
Replacing in \eq\eqref{Gamma-giant1} and recalling \eq\eqref{speed}, we thus end up with 
\begin{equation}\label{Gamma-giant}
\tilde\Gamma_\tau=
\frac{1}{\cal N}\left|
\sum_{\ell=1}^{\mathcal N}
e^{i(\tau {\bf K}\cdot {\bf R}_\ell-\phi_\ell)}
\right|^2\Gamma_\tau({{\bf R}_1})\,,
\end{equation}
with $\Gamma_\tau({{\bf R}_1})$ the decay rate of a normal atom coupled to resonator $a_{{
\bf R}_1}$ [\cf\eq\eqref{Gamma-tau}].
This vanishes precisely when \eq\eqref{condK-center}, namely the decoupling condition from valley $\tau$, holds. Therefore, when the giant atom decouples from valley $\tau$, it cannot emit into the corresponding valley-polarized edge-mode band. This result, combined with Section~\ref{sec-sel}, shows that in the present system chiral emission is achieved through selective coupling to the $K$ or $K'$ valley.

For two coupling points with strengths $g_1=g/\sqrt{2}$ and $g_2=g_1 e^{i \phi}$ as in \ref{fig:chiral}(b)-(c), \eq\eqref{Gamma-giant} reduces to
\begin{equation}
    \tilde\Gamma_\tau=\frac{1}{2}\left|
1+e^{i[\tau {\bf K}\cdot ({\bf R}_2-{\bf R}_1)-\phi]}
\right|^2\Gamma_\tau({\bf R}_1)\,,
\end{equation}
which, for ${\bf R}_2-{\bf R}_1={\bf e}_2$ and $\phi=-\pi/3$ ($\phi=\pi/3$), correctly reproduces the decay rate in \ref{fig:chiral}(b) [(c)].

\section{Conclusions}\label{sec-concl}

In this work, we have shown that giant atoms provide a natural and powerful route to implement \emph{valley-selective} light--matter interactions in synthetic honeycomb photonic lattices. By exploiting the non-local character of the emitter--bath coupling, interference among different coupling points can be engineered so as to suppress the coupling to one valley while preserving it to the other, a task unachievable for a normal atom locally coupled to a single lattice resonator.

We first analyzed a homogeneous honeycomb lattice with sublattice detuning, which breaks inversion symmetry while preserving time-reversal symmetry. In this setting, we derived the condition under which a giant atom selectively couples to only one of the two inequivalent valleys, $K$ or $K'$. We showed that this can already be achieved with only two coupling points, provided that their relative phase is appropriately tuned. Numerical simulations of spontaneous emission in the bulk confirmed these analytical predictions, showing that the emitted radiation becomes strongly valley polarized in momentum space. In this regime, the giant atom acts as a source of valley-polarized photons, while a normal atom populates both valleys symmetrically.

We then turned to the case of a spatially inhomogeneous sublattice detuning, generating a domain wall between two regions with opposite valley topology. As is well known, such a configuration supports edge modes associated with the two valleys, localized at the interface and counter-propagating along it. By combining this valley-polarized photonic environment with the selective valley coupling enabled by the giant atom, we demonstrated that the emitter undergoes chiral spontaneous emission: depending on the engineered relative phases between the coupling-point strengths, the photon is emitted preferentially into one propagation direction along the domain wall. In contrast, a normal atom, which couples equally to both valleys, emits the photon along the domain wall, with equal probability in the two opposite directions.

Our results therefore identify a physical mechanism through which non-local light--matter coupling promotes valley selectivity into directional quantum emission thus opening new perspectives for quantum-optical applications and phenomena based on the valley degree of freedom of photons. 

Importantly, the emergence of chirality here does not rely on breaking time-reversal symmetry in the photonic bath, but only in the phase-patterned giant-atom coupling to the field. This distinguishes the present setting from chiral quantum-optical platforms based on explicitly non-reciprocal or magnetic photonic metamaterials, and makes it especially appealing for implementations in experimentally accessible architectures such as circuit QED, where realizing large photonic lattices with broken time-reversal symmetry is challenging.

More broadly, our work establishes giant atoms as a versatile tool for photonic valleytronics. Beyond the single-emitter scenario considered here, the framework developed in this paper opens several promising directions. For instance, arrays of giant atoms selectively coupled to different valleys may give rise to direction-dependent collective decay, dissipative state preparation protocols, or effective interactions mediated by valley-polarized edge channels. 

In summary, we have shown that a giant atom coupled to a photonic graphene lattice can be used to engineer selective coupling to a single valley and, in the presence of valley-polarized edge modes, to realize chiral single-photon emission without requiring time-reversal-symmetry breaking in the bath. These findings introduce a new bridge between giant-atom physics, topological quantum optics, and valleytronics, and point to a flexible strategy for designing directional quantum interfaces in synthetic photonic materials. 

\acknowledgments

MAP, GLS, SC and FC acknowledge financial support from European Union-Next Generation EU through projects: Eurostart 2022
‘Topological atom-photon interactions for quantum technologies’; PRIN 2022–PNRR No. P202253RLY
‘Harnessing topological phases for quantum technologies’; THENCE–Partenariato Esteso
NQSTI–PE00000023–Spoke 2 ‘Taming and harnessing decoherence in complex networks’.  AGT acknowledges support from the CSIC Research Platform on Quantum Technologies PTI-001, Spanish project Proyecto PID2024-162384NB-I00 financiado por MICIU/AEI/10.13039/501100011033 y por FEDER,UE, from the
QUANTERA project MOLAR with reference PCI2024153449 and funded MICIU/AEI/10.13039/501100011033 and by the European Union, the Programa Fundamentos FBBVA through the grant EIC24-1-17304

\appendix

\section{Derivation of the edge modes}\label{appA}

The derivation of the approximate valley-polarized edge modes \eqref{edge-state} proceeds analogously to the standard low-energy calculation of edge modes in two-dimensional topological lattices \cite{Asboth2016ShortCourse}.

In the presence of a non-uniform sublattice detuning $\Delta(x)$ that varies slowly on the scale of the unit cell, the low-energy effective Hamiltonian in each valley reads [\cf\eq\eqref{htau}]
\begin{equation}\label{Htaux}
H_\tau = v\left(\tau p_x \sigma_x + p_y \sigma_y\right) + \Delta(x)\sigma_z\,.
\end{equation}
Under periodic boundary conditions along the $y$ axis, $k_y$ is a good quantum number. Accordingly, for each value of $k_y$ one can diagonalize the Hamiltonian $H_\tau(k_y)$ obtained from \eqref{Htaux} under the replacement $p_y\rightarrow k_y$.

We search for eigenstates of $H_\tau(k_y)$ whose real-space wavefunction has the form $\Psi_\tau(x,y)=e^{ik_y y}\,\psi_\tau(x)$ with $\psi_\tau(x)$ the $x$-dependent spinor defined by 
$\psi_\tau(x)=\left(\psi_a(x), \psi_b(x)\right)^T$. 
Enforcing the Schr\"odinger equation $H_\tau(k_y) \Psi_\tau(x,y)=E \Psi_\tau(x,y)$ yields
\begin{equation}
\left[
v\left(-i\tau \sigma_x \partial_x + k_y \sigma_y\right)
+\Delta(x)\sigma_z
\right]\psi_\tau(x)
=
E\,\psi_\tau(x)\,,
\label{eq: schrodinger eq x}
\end{equation}
where we expressed $p_x$ in real space. \eq\eqref{eq: schrodinger eq x} is equivalent to the system of differential equations
\begin{equation}\label{system}
\begin{aligned}
\Delta(x)\psi_a - i\tau v\, \psi_b' - i v k_y \psi_b &= E \psi_a\,, \\
-\Delta(x)\psi_b - i\tau v\, \psi_a' + i v k_y \psi_a &= E \psi_b\,.
\end{aligned}
\end{equation}
where $(\cdot)' \equiv \partial_x(\cdot)$.

Focusing now on the case $k_y{=}0$, we search for a zero-energy solution of \eqref{system}. The differential system thus simplifies to a system of coupled first-order linear differential equations with $x$-dependent coefficients
\begin{align}
\Delta(x)\psi_a - i\tau v\, \psi_b' &= 0, \\
-\Delta(x)\psi_b - i\tau v\, \psi_a' &= 0.
\end{align}
Under the boundary conditions $\psi_a(\pm \infty)=\psi_b(\pm \infty)=0$, the solution is 
\begin{equation}\label{psi1}
\psi_\tau(x) = \mathcal N\, \exp\left[ -\frac{1}{v}\int_0^x \Delta(x')\,dx' \right]
\begin{pmatrix}
1\\
i\tau
\end{pmatrix}.
\end{equation}
The normalization condition of \eqref{psi1} is easily seen to reduce to the equation
\begin{equation}\label{norm}
  4|\mathcal{N}|^2 \lambda\int \limits_{0}^{\infty}d\xi \cosh^{-\beta}\left(\xi\right) = 1 \,,
\end{equation}
where $\xi=x/\lambda$ and $\beta = 2\Delta_0\lambda/v$. The integral can be calculated as
\begin{equation}
\int \limits_{0}^{\infty}d\xi \cosh^{-\beta}\left(\xi\right) = \frac{\sqrt{\pi}}{2}\frac{\Gamma \left(\frac{\beta}{2} \right)}{\Gamma \left(\frac{\beta+1}{2} \right)}
\end{equation}
Solving \eq\eqref{norm} for ${\cal N}$ and replacing in \eq\eqref{psi1}, we finally end up with \eq\eqref{psix}.

To derive the dispersion relation, we now consider a finite $k_y$,
\begin{equation}
H_\tau(k_y)=H_\tau(0)+v_F k_y \sigma_y.
\end{equation}
Since the bound state satisfies $\sigma_y \psi_\tau = \tau \psi_\tau$ we can immediately obtain
\begin{equation}
E_\tau(k_y)=\tau v_F k_y.
\end{equation}
	
\bibliography{PSCGTDBC_v1}
	
\appendix	
\end{document}

%% file: preamble.tex
\usepackage[english]{babel}
\usepackage[utf8]{inputenc}
\usepackage{dcolumn}
\usepackage{bm}
\usepackage{graphicx}
\usepackage{amsmath,amsfonts,amssymb}
\usepackage{physics}
\usepackage{enumitem}
\usepackage{latexsym}
\usepackage{array}
\usepackage{epsfig}
\usepackage{color}
\usepackage[
    colorlinks=true,
    linkcolor=blue,
    urlcolor=blue,
    citecolor=blue,
    pdfusetitle]{hyperref}
\usepackage[T1]{fontenc}
\usepackage[sc]{mathpazo}
\usepackage{orcidlink}
\usepackage{bibunits}

\usepackage{varioref}
\usepackage{comment}
\labelformat{figure}{Fig.\;#1}

\newcommand{\aref}[1]{\hyperref[#1]{Appendix~\ref*{#1}}}

\newcommand{\eq}{Eq.\,}
\newcommand{\eqs}{Eqs.\,}

\newcommand{\cf} {cf.~}

\newcommand{\rrefs} {Refs.\,}
\DeclareMathOperator{\Hc}{H.c.}

\newcommand{\R} {{\bf R}}